\documentclass[]{raa}            
\usepackage{graphicx}
\usepackage{natbib}
\usepackage{hyperref,ifthen}
\usepackage{multirow}
\usepackage{amssymb,amsmath}

\providecommand{\bm}[1]{\mbox{\boldmath$#1$\unboldmath}}

\begin{document}

\newcommand{\beq}[1]{\begin{equation}\label{#1}}
 \newcommand{\eeq}{\end{equation}}
 \newcommand{\bea}{\begin{eqnarray}}
 \newcommand{\eea}{\end{eqnarray}}
 \def\disp{\displaystyle}

   \title{Comparison of cosmological models using standard rulers and candles
}

 \volnopage{ {\bf 2015} Vol.\ {\bf XX} No. {\bf XX}, 000--000}
   \setcounter{page}{1}
\author{Xiaolei Li \inst{1}, Shuo Cao \inst{1*}, Xiaogang Zheng \inst{1}, Song Li \inst{2}, and Marek Biesiada \inst{1,3}}

\institute{Department of Astronomy, Beijing Normal University, Beijing 100875, China; {\it caoshuo@bnu.edu.cn}\\
     \and
   Department of Physics, Capital Normal University, Beijing 100048, China\\
\and
Department of Astrophysics and Cosmology, Institute of Physics, University of Silesia, Uniwersytecka 4, 40-007 Katowice, Poland\\
%
\vs \no
   {\small Received [year] [month] [day]; accepted [year] [month] [day] }
}

\abstract{ In this paper, we used standard rulers and standard
candles (separately and jointly) to explore five popular dark energy
models under assumption of spatial flatness of the Universe. As
standard rulers, we used a data set comprising 118 galactic-scale
strong lensing systems (individual standard rulers if properly
calibrated for the mass density profile) combined with BAO
diagnostics (statistical standard ruler). Supernovae Ia served as
standard candles. Unlike in the most of previous statistical studies
involving strong lensing systems, we relaxed the assumption of
singular isothermal sphere (SIS) in favor of its generalization: the
power-law mass density profile. Therefore, along with cosmological
model parameters we fitted the power law index and its first
derivative with respect to the redshift (thus allowing for mass
density profile evolution). It turned out that the best fitted
$\gamma$ parameters are in agreement with each other irrespective of
the cosmological model considered. This demonstrates that galactic
strong lensing systems may provide a complementary probe to test the
properties of dark energy. Fits for cosmological model parameters
which we obtained are in agreement with alternative studies
performed by the others. Because standard rulers and standard
candles have different parameter degeneracies, combination of
standard rulers and standard candles gives much more restrictive
results for cosmological parameters. At last, we attempted at model
selection based on information theoretic criteria (AIC and BIC). Our
results support the claim, that cosmological constant model is still
the best one and there is no (at least statistical) reason to prefer
any other more complex model. }

\authorrunning{ Xiaolei Li, et al.}
\titlerunning{Comparison of cosmological models using standard rulers and candles}
   \maketitle


\section{Introduction}
\label{sec:introduction} One of the most important issues in modern
cosmology is the accelerated expansion of the universe, deduced from
Type Ia supernovae \citep{Riess98, Perlmutter99} and also confirmed
by other independent probes, such as Cosmic Microwave Background
(CMB)\citep{Pope04} and the Large Scale Structure (LSS)
\citep{Spergel03}. In order to explain this phenomenon, a new
component, called dark energy, which fuels the cosmic acceleration
due to its negative pressure and may dominate the universe at late
times has been introduced.

Although cosmological constant $\Lambda$ \citep{Peebles03}, the
simplest candidate for dark energy, seems to fit in with current
observations, yet it suffers from the well-known fine tuning and
coincidence problems. Therefore a variety of dark energy models,
including different dark energy equation of state (EoS)
parametrizations such as XCDM model \citep{Ratra98}, and
Chevalier-Polarski-Linder (CPL) model \citep{Chevalier01,Linder03}
have been put forward, each of which has its own advantages and
problems in explaining the acceleration of the universe. Yet, the
nature of dark energy still remains unknown. It might also be
possible that observed accelerated expansion of the Universe is due
to departures of the true theory of gravity from General Relativity,
e.g. due to quantum nature of gravity or possible
multidimensionality of the world. Hence such models like
Dvali-Gabadadze-Porrati -- inspired by brane theory or Ricci dark
energy inspired by the holographic principle have been proposed.
Having no clear preference from the side of theory and in order to
learn more about dark energy, we have to turn to the sequential
upgrading of observational fits of quantities which parametrize the
unknown properties of dark energy (such as density parameters or
coefficients in the cosmic equation of state) and seeking coherence
among alternative tests. In this paper we highlight the usefulness
of strong lensing systems to assess the parameters of several
popular dark energy models. Because strong lensing systems are
sensitive to angular diameter distance we supplement our analysis
with Baryon Acoustic Oscillations (BAO) data and compare our results
with the inference made using luminosity distances measured with SN
Ia.

Strong gravitational lensing has recently developed into a serious
technique in extragalactic astronomy (galactic structure studies)
and in cosmology. First of all, the angular separation between
images (determined by the Einstein radius of the lens) can be used
to constrain and model the mass distribution of lens
\citep{Harvard-Smithsonian96}. Secondly, time delay between images
are additional sources of constraints on the mass distribution of
the lens. Strong lensing time delays have recently developed into a
promising new technique to constrain cosmological parameters -- the
Hubble constant in the first place \citep{Suyu09}. Finally, strong
lensing systems are becoming an important tool in cosmology. Earlier
attempts to use such systems for constraining cosmological
parameters were based on comparison between theoretical (depending
on the cosmological model) and empirical distributions of image
separations \citep{Chae02,Cao12a} or lens redshifts
\citep{Ofek03,Cao12c} in observed samples of lenses. Another
approach is based on the fact that image separations in the system
depend on angular diameter distances to the lens and to the source,
which in turn are determined by background cosmology. This method
applied in the context of dark energy was first proposed in the
papers of \citet{Futamase01,Biesiada06,Gilmore09} and further
developed in recent works \citep{Biesiada10,Cao12b}.

However, there are two well known challenges to this method as a
cosmological probe. The first issue is that detailed mass
distribution of the lens and its possible evolution in time are not
clear enough. The second challenge is the limited number of observed
gravitational lensing systems with complete spectroscopic and
astrometric information necessary for this technique. It is only
quite recently when reasonable catalogs of strong lensing systems
are becoming available. In this paper, we use an approach proposed
by \citet{Cao15} where the lensing galaxy is assumed to have
spherically symmetric mass distribution described by the power-law
slope factor which is allowed to evolve with redshift.

Baryon Acoustic Oscillations (BAO) and strong lensing systems
together, constitute such independent standard rulers which may have
different degeneracies in the parameter space of dark energy models
\citep{Marek11}. Moreover, we also take supernovae Ia for comparison
as an independent probe (standard candles). In Section~\ref{models},
we present cosmological models considered and the corresponding
results. In order to compare dark energy models with different
numbers of parameters and decide which model is preferred by the
current data, in Section~\ref{AIC} we apply two model selection
techniques i.e. the Akaike Criterion (AIC) and Bayesian Information
Criterion (BIC). Finally the results are summarized in
Section~\ref{conclusions}.

\section{Method and data} \label{data}

\subsection{Strong lensing systems}

Strong lensing system with the intervening galaxy acting as a lens
usually produces multiple images of the source. Image separation
depends in the first place on the mass of the lens (suitably
parametrized by stellar velocity dispersion) but also on the angular
diameter distances between the lens and the source and between the
observer and the lens. Angular diameter distance between two objects
at redshifts $z_1$ and $z_2$ respectively, is determined by
background cosmology:
\begin{equation}
{D(z_1,z_2;\bf{p})}= {\frac{c} {{H_0}({1+z_2)}}}{\int_{z_1}^{z_2} {\frac{dz}{E(z;\bf{p})}} }
\end{equation}
where $E(z;\mathbf{p}) = H(z;\mathbf{p})/ H_0$ is the dimensionless
expansion rate, $H_0$ is the Hubble constant and $\bf{p}$ denotes
the parameters of a particular cosmological model considered.

Under assumption of the singular isothermal sphere (SIS) model,
which so far was a standard one for elliptical galaxies acting as
lenses, the Einstein radius $\theta_E$ is given by
\begin{equation} \label{thetaE_SIS}
{{\theta}_E}=4{\pi}{\frac{D_{ls}} {D_s}}{\frac{{\sigma}^2_{SIS}} {c^2}}
\end{equation}
where $D_{ls}$ and $D_s$ are angular diameter distances between
lensing galaxy and the source and between the observer and the
source, respectively. If the Einstein radius and the velocity
dispersion of lensing galaxy are known from observations, the ratio
of angular diameter distances $D_{ls}/D_s $ can be obtained from
Eq.~(\ref{thetaE_SIS}). The main challenge here is how to get the
velocity dispersion of lensing galaxy $\sigma_{SIS}$ (which is the
SIS model parameter) from central stellar velocity dispersion
$\sigma_0$ obtained from spectroscopy. Previous analysis using
strong gravitational lensing systems, took the phenomenological
approach to relate these two velocity dispersions through
$\sigma_{SIS}=f_E\sigma_0$, where $f_E$ was a free parameter with
$0.8<f_E^2<1.2$ \citep{Ofek03,Cao12b}.

In this paper, we generalize the SIS model to spherically symmetric
power-law mass distribution $\rho \sim r^{-\gamma}$ \citep{Cao15}.
Accordingly, the Einstein radius can be rewritten as
\begin{equation}
{{\theta}_E}=4{\pi}{\frac{D_{ls}} {D_s}}{\frac{{\sigma}^2_{ap}} {c^2}}\left(\frac{{\theta}_E}{{\theta}_{ap}}\right)^{2-\gamma}f(\gamma)
\end{equation}
where $\sigma_{ap}$ is the velocity dispersion inside the aperture of size ${\theta}_{ap}$,
and
\begin{equation}
{f(\gamma)}=-{\frac{1}{\sqrt{\pi}}}{\frac{(5-2{\gamma})(1-{\gamma})}{3-{\gamma}}}{\frac{\Gamma(\gamma-1)}{\Gamma(\gamma-{\frac{2}{3})}}}\left[\frac{(\Gamma(\frac{\gamma}{2})-\frac{1}{2})}{\Gamma(\frac{\gamma}{2})}\right]^2
\end{equation}
As a result, observational value of the angular diameter
distance ratio reads:
\begin{equation}
{D^{obs}}={\frac{D_{ls}}{{D_{s}}}}={\frac{{c^2}{\theta_E}}{4{\pi}{{\sigma}^2_{ap}}}}\left(\frac{\theta_{ap}}{\theta_E}\right)^{2-\gamma}{f^{-1}(\gamma)}
\end{equation}
and its theoretical counterpart can be obtained from Eq.~(1)
\begin{equation}
{D^{th}(z_l,z_s;\bf{p})}={\frac{D_{ls}{\bf{(p)}}}{D_s\bf{(p)}}}={\frac{{\int_{z_l}^{z_s} {\frac{dz}{E(z;\bf{p})}} }}{{\int_{0}^{z_s} {\frac{dz}{E(z;\bf{p})}} }}}
\end{equation}

Then one can constrain cosmological models by minimizing the
$\chi^2$ function given by
\begin{equation}
 {\chi^2_{SL}\bf{(p)}}=\sum^N_{i=1} \left[ \frac{{D^{th}(z_{l,i},z_{s,i};\bf{p})}-{D^{obs}(\sigma_{ap,i},\theta_{E,i};\gamma)}}{\sigma_{D,i}} \right]^2
\end{equation}
where the variance of $D^{obs}$ is
\begin{equation}
 {{\sigma^2_{D,i}}} = D^{obs}(\sigma_{ap,i},\theta_{E,i};\gamma)^2 \left[ 4 \left(  \frac{\sigma_{\sigma_{ap}}}{\sigma_{ap}}\right)^2+{(1-\gamma)}^2 \left( \frac{\sigma_{\theta_E}}{\theta_E} \right)^2 \right].
\end{equation}

In order to calculate $\sigma_D$, we assumed the fractional
uncertainty of the Einstein radius at the level of $5\%$ (i.e.,
$\frac{\sigma_{\theta_E}} {\theta_E}$=0.05) for all the lenses,
uncertainties of the velocity dispersion were taken from the data
set --- see \citep{Cao15} for details.

We treated the mass density power-law index of lensing galaxies as a
free parameter to be estimated together with cosmological
parameters. Moreover, since it has recently been claimed that
$\gamma$ index of elliptical galaxies might evolved with redshift
\citep{Ruff11}, we assumed the linear relation for $\gamma$:
$\gamma=\gamma_0+\gamma_1z_l$. Furthermore, when dealing with a
sample of lenses instead of a single lens system, we followed the
standard practice and transformed velocity dispersion measured
within actual circular aperture to the one within circular aperture
of radius $R_{eff}/2$ (half the effective radius) according to the
prescription of \citep{Jorgensen95}:
$\sigma_0=\sigma_{ap}{(\theta_{eff}/(2\theta_{ap}))}^{-0.04}$. Such
procedure has an advantage of standardizing measured velocity
dispersions within the sample and introduces negligible terms to the
error budget --- for details see \citet{Cao15}.

In this paper, we use a combined sample of $n=118$ strong lensing
systems from SLACS (57 lenses taken from \citet{Bolton08,Auger09}),
BELLS (25 lenses taken from \citet{Brownstein12}), LSD (5 lenses
from \citet{Treu02,Koopmans02,Treu04}) and SL2S (31 lenses taken
from \citet{Sonnenfeld13a,Sonnenfeld13b}), which is the most recent
compilation of galactic scale strong lensing data. This sample is
compiled and summarized in Table 1 of \citet{Cao15}, in which all
relevant information necessary to perform cosmological model fit can
be found.

\subsection{Baryon Acoustic Oscillations}

Baryon Acoustic Oscillations (BAO) refer to regular, periodic
fluctuations in the density of visible baryonic matter in the
Universe (the large scale structure). Being ``the statistical
standard ruler'' they are commonly used to investigate dark energy.
From BAO observations, we used the BAO distance ratio
$r_{s}(z_{d})/D_{V}(z)$ measured by the Sloan Digital Sky Survey
(SDSS) data release 7 (DR7) \citep{Padmanabhan12}, SDSS-III Baryon
Oscillation Spectroscopic Survey (BOSS) \citep{Anderson12}, the
clustering of WiggleZ survey \citep{Blake12} and 6dFGS survey
\citep{Beutler11}:

\begin{equation} \label{BAO distance}
d_{z}=\frac{r_{s}(z_{d})}{D_{V}(z)}.
\end{equation}

The meaning of the quantities $r_s(z_d)$ and $D_V(z)$ is explained below. The effective distance is given by
\begin{equation}
D_{V}(z)\equiv[(1+z)^{2}D^{2}_{A}(z)\frac{cz}{H(z)}]^{1/3}
\end{equation}
where $D_{A}(z)$ is the angular diameter distance and $H(z)$ is the Hubble parameter. The comoving sound horizon scale at the baryon drag epoch is
\begin{equation}
r_{s}(z_{d})=\int^{\infty}_{z_{d}}\frac{c_{s}(z')dz'}{E(z')}
\end{equation}
where the sound speed is given by the formula:
$c_{s}(z)=1/\sqrt{3[1+\bar{R_{b}}/(1+z)]}$ where
$\bar{R_{b}}=3\Omega_{b}h^{2}/(4\times2.469\times10^{-5})$, and the
drag epoch redshift is fitted as
\begin{equation}
z_d=\frac{1291(\Omega_{m}h^{2})^{0.251}}{1+0.659(\Omega_{m}h^{2})^{0.828}}[1+b_{1}(\Omega_{b}h^{2})^{b_{2}}]
\end{equation}
where $b_{1}=0.313(\Omega_{m}h^{2})^{-0.419}[1+0.607(\Omega_{m}h^{2})^{0.674}]$ and
$b_{2}=0.238(\Omega_{m}h^{2})^{0.233}$.

The $\chi^{2}$ function for BAO data is defined as
\begin{equation}
\chi^{2}_{BAO}=\bf(x-d)^{T}(C_{BAO}^{-1})(x-d),
\end{equation}
where
\begin{equation}
\begin{aligned}
{\bf{x-d}}=[r_{s}/D_{V}(0.1)-0.336, D_{V}(0.35)/r_{s}-8.88,\\
 D_{V}(0.57)/r_{s}-13.67,r_{s}/D_{V}(0.44)-0.0916,\\
r_{s}/D_{V}(0.60)-0.0726, r_{s}/D_{V}(0.73)-0.0592]
\end{aligned}
\end{equation}
and $C^{-1}_{BAO}$ is the corresponding inverse covariance matrix
taken after \citet{Hinshaw13}. In order to constrain cosmological
parameters with standard rulers, i.e. strong lensing systems and BAO
we used the joint $\chi^2$  defined as:
 \begin{equation}
\chi^2_{SL+BAO}={\chi^{2}_{SL}}+{\chi^{2}_{BAO}}
\end{equation}

\subsection{Supernovae Ia}

Up to now, we discussed standard rulers, but it were standard
candles (SN Ia) which kicked off the story of accelerated expansion
of the Universe. Since then they remained a reference point for
discussions and tests of cosmological models. Because standard
rulers (SL+BAO) and standard candles measure the distances based on
different concepts, cosmological inferences based on them have
different degeneracies in parameter space. Therefore, we also
considered constraints on cosmologies coming from SN Ia
observations: for comparison and also for the sake of
complementarity. In this paper, we used the latest Union2.1
compilation released by the Supernova Cosmology Project
Collaboration consisting of 580 SN Ia data points \citep{Suzuki12},
taking into consideration systematic errors of the observed distance
modulus \citep{Cao14}. The Hubble constant $H_0$ was treated as a
nuisance parameter and was marginalized over with a flat prior. The
$\chi^2_{SN}$ function for the supernovae data is given by
\begin{eqnarray}
\chi_{SN}^2&=&\sum_{i,j}\alpha_iC_{SN}^{-1}(z_i,z_j)\alpha_j \nonumber\\
           &-&\frac{[\sum_{ij}\alpha_iC_{SN}^{-1}(z_i,z_j)-\ln10/5]^2}{\sum_{ij}C_{SN}^{-1}(z_i,z_j)}\nonumber\\
           &-&2\ln\bigg(\frac{\ln10}{5}\sqrt{\frac{2\pi}{\sum_{ij}C_{SN}^{-1}(z_i,z_j)}}\bigg),
 \label{chiSN2}
\end{eqnarray}
where $\alpha_i=\mu_{obs}(z_i)-25-5\log_{10}[H_0D_L(z_i)/c]$ and
$C_{SN}(z_i,z_j)$ is the covariance matrix.

Finally, we also performed a joint analysis with both standard
rulers and standard candles using the combined chi-square function:
\begin{equation}
\chi^2_{tot}={\chi^{2}_{SL}}+{\chi^{2}_{BAO}} + {\chi^{2}_{SN}}
\end{equation}

\section{Cosmological models and results} \label{models}

In this section, we choose several popular dark energy models and
estimate their best fitted parameters using the standard rulers
(strong lensing systems and BAO), standard candles and using all
these cosmological probes jointly. We also examine consistency of
our findings with other independent results from the literature.
Table~\ref{tab:hubble} reports the Hubble function for different
cosmological models considered. Throughout our paper we report the
best fit values, and corresponding 1$\sigma$ uncertainties ($68 \%$
confidence intervals) for each class of models considered.
Moreover,we assume spatially flat Universe. The results of
cosmological parameters from standard rulers (SL+BAO) are presented
in Table~\ref{tab:result}.

 \begin{table*}
\caption{\label{tab:hubble} Hubble function for different cosmological models considered.}

\begin{center}
\begin{tabular}{c|c|cc}
\hline \hline
Model & Hubble function & Cosmological parameters\\
\hline $\Lambda$CDM & ${H^2(z)}={{H^2_0}{[\Omega_m(1+z)^3+(1-\Omega_m)]}}$ & $\Omega_m$\\
\hline XCDM & ${H^2(z)}={{H^2_0}{[\Omega_m(1+z)^3+\Omega_{\Lambda}(1+z)^{3(1+w)}]}}$ & $\Omega_m$, $w$\\
\hline CPL & ${H^2(z)}={{H^2_0}{[\Omega_m(1+z)^3+\Omega_{\Lambda}(1+z)^{3(1+w_0+w_1)}{exp(\frac{-3w_1z}{1+z})}]}}$ & $w_0$, $w_1$\\
\hline RDE & ${H^2(z)}={{H^2_0}}[\frac{2\Omega_m}{2-\beta}(1+z)^3+(1-\frac{2\Omega_m}{2-\beta})(1+z)^{4-2/\beta}]$ & $\Omega_m$, $\beta$ \\
\hline DGP & ${H^2(z)}={{H^2_0}}{[({\sqrt{\Omega_m(1+z)^3+\Omega_{rc}}}+{\sqrt{\Omega_{rc}}})^2]}$ & $\Omega_m$ \\
\hline \hline
\end{tabular}
\end{center}
\end{table*}

\begin{table*}
\caption{\label{tab:result} Best fits for different cosmological models from standard rulers (SL+BAO)}
\begin{center}
\begin{tabular}{c|c|cc}
\hline \hline
Model & Cosmological parameters & Mass density slope parameters \\
\hline $\Lambda$CDM & $\Omega_m=0.279^{+0.022}_{-0.022}$ & $\gamma_0=2.094^{+0.053}_{-0.056}$,$\gamma_1=-0.053^{+0.103}_{-0.102}$  \\
\hline XCDM & ${\Omega_m}=0.282^{+0.021}_{-0.023}$,  $w=-0.917^{+0.194}_{-0.188}$  & $\gamma_0=2.088^{+0.055}_{-0.056}$,$\gamma_1=-0.054^{+0.104}_{-0.102}$\\
\hline CPL &  $w_0=-0.879^{+0.325}_{-0.314}$, $w_1=-0.464^{+0.870}_{-0.710}$   & $\gamma_0=2.087^{+0.055}_{-0.056}$,$\gamma_1=-0.055^{+0.105}_{-0.105}$ \\
\hline RDE &  ${\Omega_m}=0.201^{+0.017}_{-0.019}$,  $\beta=0.566^{+0.087}_{-0.086}$ & $\gamma_0=2.087^{+0.052}_{-0.054}$,$\gamma_1=-0.052^{+0.104}_{-0.102}$ \\
\hline DGP & $\Omega_m=0.269^{+0.014}_{-0.016}$ & $\gamma_0=2.074^{+0.050}_{-0.051}$, $\gamma_1=-0.047^{+0.101}_{-0.102}$  \\
\hline \hline
\end{tabular}
\end{center}
\end{table*}

\subsection{Standard cosmological model}

Currently standard cosmological model, also known as the
$\Lambda$CDM model is the simplest one with constant dark energy
density present in the form of cosmological constant $\Lambda$. It
agrees very well with various observational data such as CMB
anisotropies, and LSS distribution \citep{Pope04,Riess98}, etc.
Formally, one can say, that cosmic equation of state here is simply
$w=p/{\rho}=-1$.

\begin{figure}[h]
\begin{center}
\includegraphics[scale=0.5]{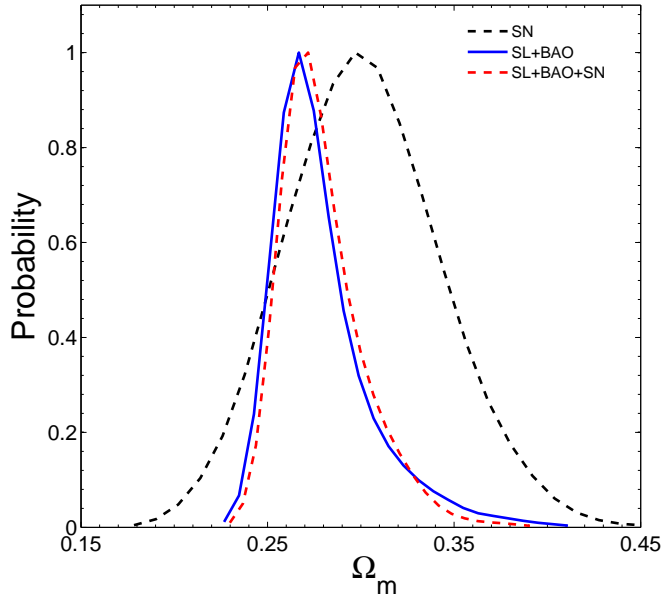}
\end{center}
\caption{\label{tab:LCDM}Constraints on $\Lambda$CDM model. The blue curve is the result from SL+BAO, the black
one is from SN, and the the red one is from SL+BAO+SN. \label{LCDM}}
\end{figure}

If flatness of the FRW metric is assumed, the only cosmological
parameter of this model is ${\bf{p}}=\left\{\Omega_m\right\}$. We
obtain $\Omega_m=0.279^{+0.022}_{-0.022}$ from standard rulers
(SL+BAO), ${\Omega_m}=0.301^{+0.040}_{-0.041}$ from standard candles
(SN Ia) and ${\Omega_m}=0.280^{+0.020}_{-0.020}$ from the
combination of standard rulers and standard candles. The results are
presented in Fig.~\ref{tab:LCDM} and Table~\ref{tab:result}. One can
see that standard rulers have considerable leverage on the joint
analysis.

For comparison, it is necessary to refer to earlier results obtained
with other independent measurements. By studying peculiar velocities
of galaxies, the only method sensitive exclusively to the matter
density parameter, \citet{Feldman03} obtained
$\Omega_m=0.30^{+0.17}_{-0.17}$, a value which agrees with our joint
analysis within the 1$\sigma$ range. Based on the WMAP 9-year
results, \citet{Hinshaw13} gave the best-fit parameter:
${\Omega_m}=0.279\pm 0.025$ for the flat $\Lambda$CDM model,  which
is in perfect agreement with our standard rulers result. Let us note
that cosmological probe inferred from CMB anisotropy measured by
WMAP is also a standard ruler --- comoving size of the acoustic
horizon. It is the same ruler as in BAO technique, hence the strong
consistency revealed here could be expected. More recently, using
the corrected redshift - angular size relation for quasar sample,
\citet{Cao15b} obtained $\Omega_m=0.292^{+0.065}_{-0.090}$ in the
spatially flat $\Lambda$CDM cosmology, which is also in a very good
agreement with our findings.

\begin{figure}[h]
\begin{center}
\includegraphics[scale=0.35]{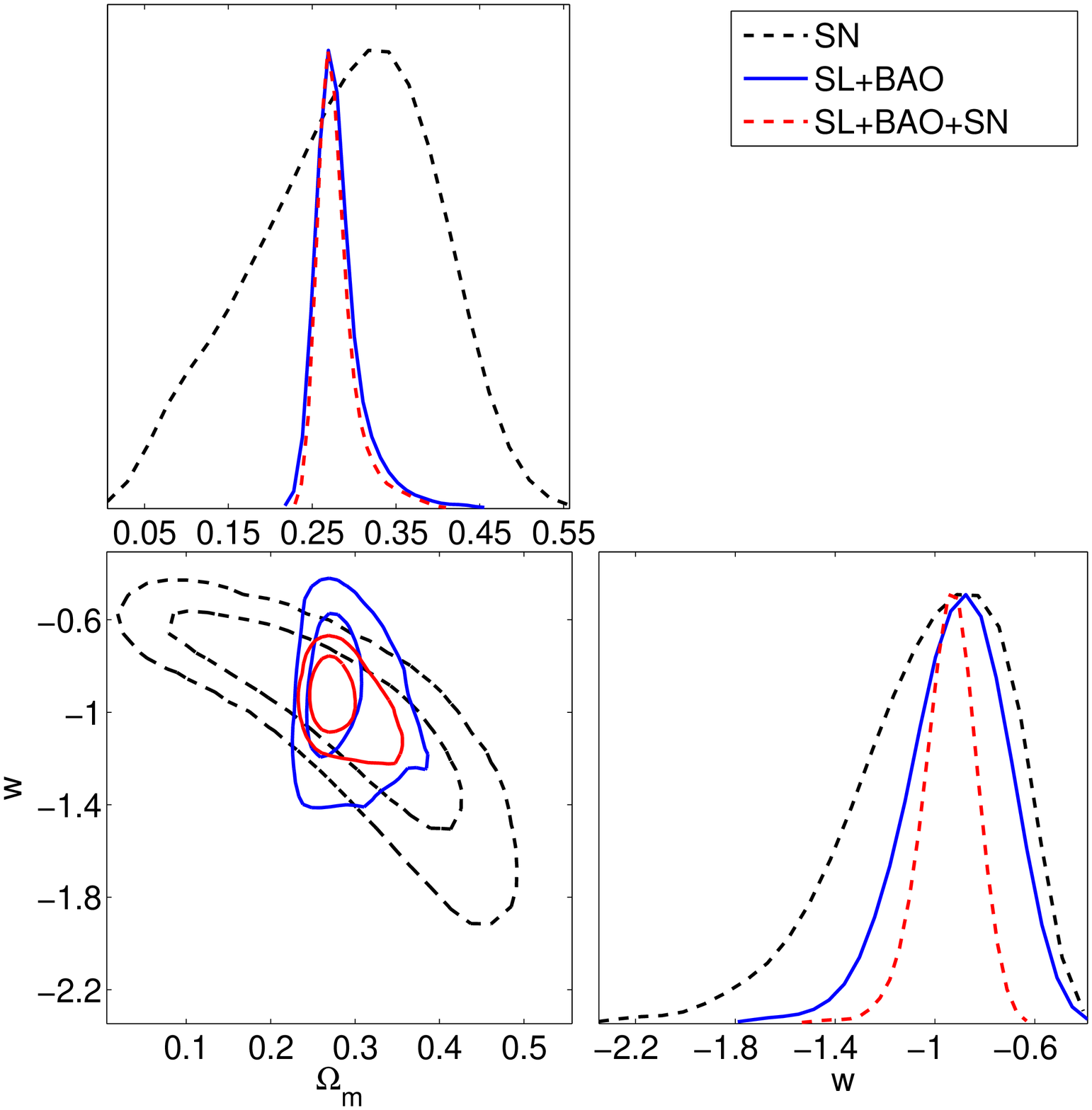}
\end{center}
\caption{\label{tab:XCDM}Constraints on XCDM model. The
blue curve is the result from SL+BAO, the black one is from SN, and
the the red one is from SL+BAO+SN. }
\end{figure}

\begin{figure}[h]
\begin{center}
\includegraphics[scale=0.35]{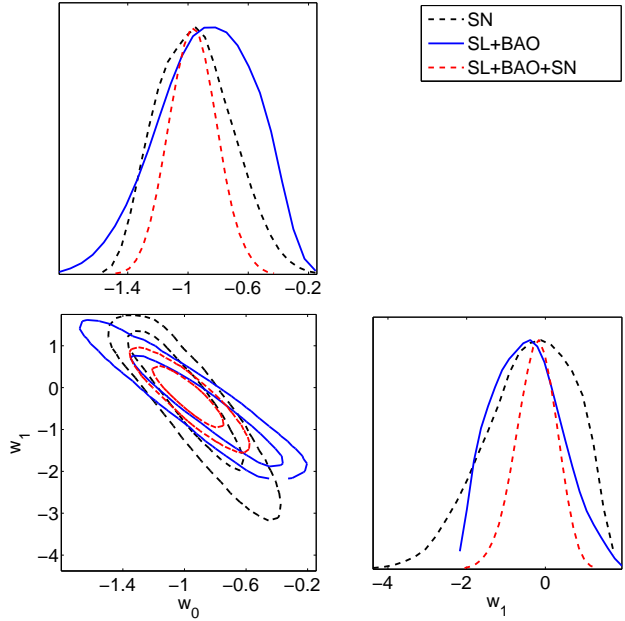}
\end{center}
\caption{ \label{tab:CPL} Constraints on CPL parameterization with fixed
$\Omega_m=0.315$. The blue curve is the
result from SL+BAO, the black one is from SN, and the the red one
is from SL+BAO+SN.\label{CPL}}
\end{figure}

\begin{figure}[h]
\begin{center}
\includegraphics[scale=0.35]{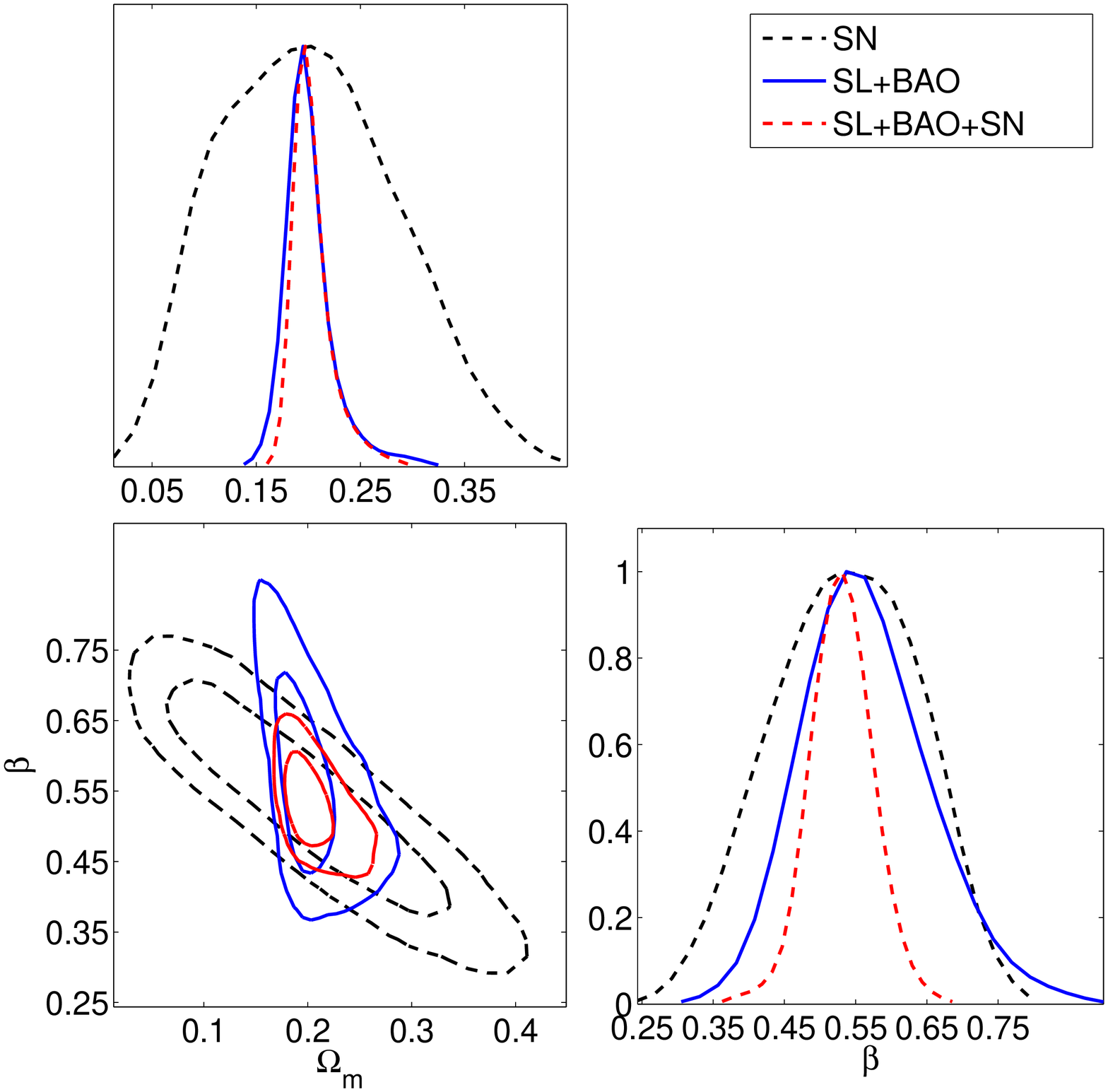}
\end{center}
\caption{ \label{tab:RDE} Constraints on RDE model. The
blue curve is the result from SL+BAO, the black one is from SN, and
the the red one is from SL+BAO+SN. \label{RDE}}
\end{figure}

\begin{figure}[h]
\begin{center}
\includegraphics[scale=0.5]{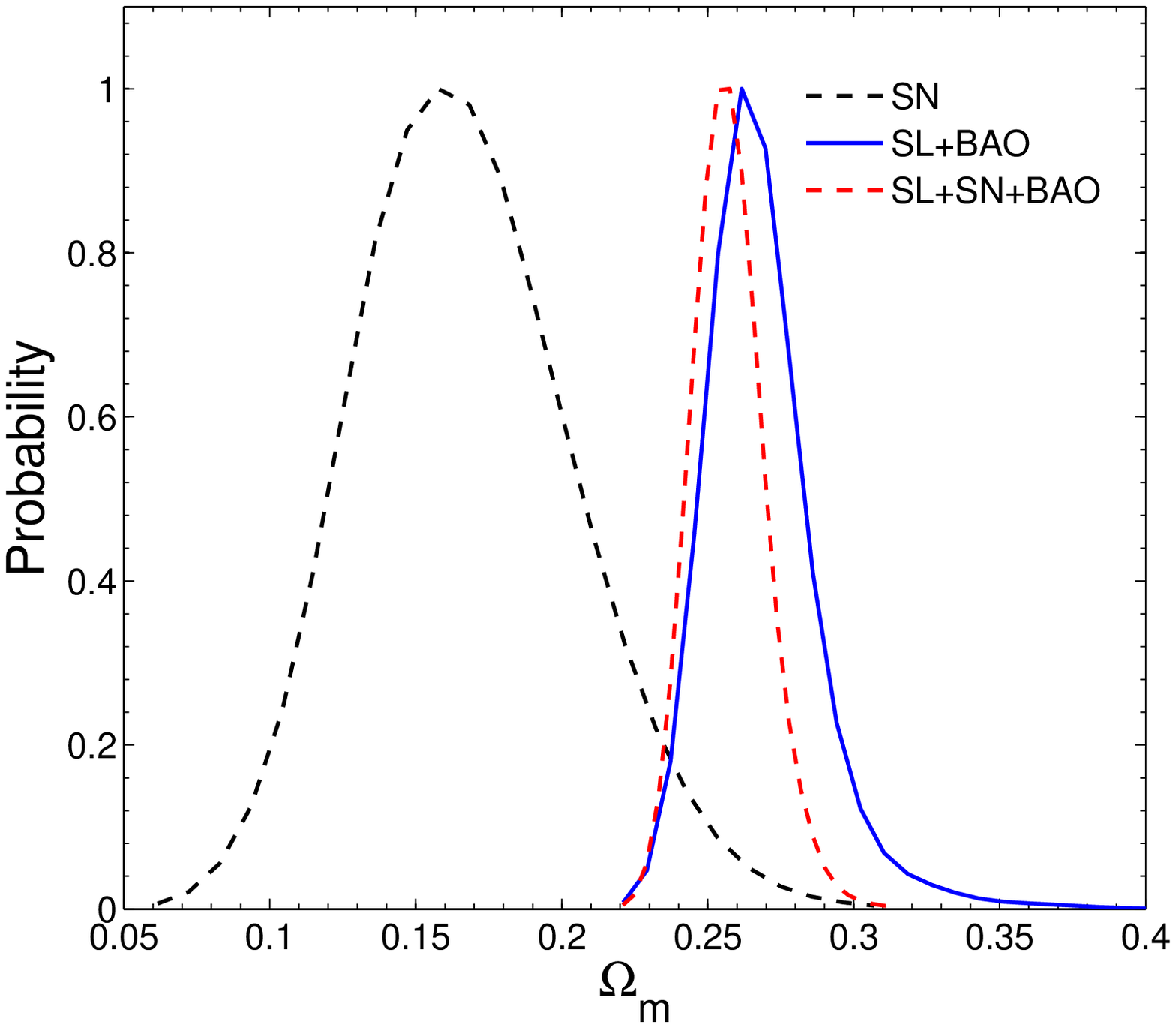}
\end{center}
\caption{\label{tab:DGP}  Constraints on DGP. The blue
curve is the result from SL+BAO, the black one is from SN, and the
the red one is from SL+BAO+SN. \label{DGP}}
\end{figure}

\subsection{Dark energy with constant equation of state}
In this case, dark energy is described by a hydrodynamical
energy-momentum tensor with constant equation of state coefficient
$w=p/{\rho}$, which leads to cosmic acceleration whenever $w<-1/3$
\citep{Ratra98}.

For standard rulers, standard candles and the combined analysis,
confidence regions (corresponding to 68.3\% and 95.8\% confidence
levels) in the ($\Omega_m$, w) plane are shown in
Fig.~\ref{tab:XCDM}, with the best-fit parameters:
$\left\{\Omega_m,w\right\}=\left\{0.282^{+0.021}_{-0.023},-0.917^{+0.194}_{-0.188}\right\}$,
$\left\{\Omega_m,w\right\}=\left\{0.287^{+0.104}_{-0.111},-0.917^{+0.308}_{-0.304}\right\}$,
and
$\left\{\Omega_m,w\right\}=\left\{0.280^{+0.018}_{-0.019},-0.947^{+0.102}_{-0.094}\right\}$
from SL+BAO, SN, and SL+BAO+SN, respectively. As can be seen from
Fig.~\ref{tab:XCDM}, standard rulers (SL+BAO) and standard candles
(SN) have different degeneracies in the parameter space.
Consequently, their restrictive power is different. This fact makes
the joint constraint more restrictive. Our results are in perfect
agreement with the previous results obtained from the ESSENCE
supernova survey team \citep{Wood-Vasey07} who obtained
$\left\{\Omega_m,w\right\}=\left\{0.274^{+0.033}_{-0.020},-1.07\pm{+0.09}\pm{-0.12}\right\}$
and the Union1 SNIa compilation \citep{Kowalski08} whose results
were following:
$\left\{\Omega_m,w\right\}=\left\{0.274^{+0.033}_{-0.020},-0.969^{+0.059}_{-0.063}{^{+0.063}_{-0.066}}\right\}$.

\subsection{Chevalier-Polarski-Linder (CPL) model}

Constant cosmic equation of state of the XCDM cosmology is only a
phenomenological description, which cannot be ultimately true. Being
fundamentally different from the cosmological constant, it must have
some dynamical reason e.g. in a scalar field settling down on the
attractor. Therefore one should expect that such scalar field was
evolving and left the trace of its evolution on the equation of
state. So, it would be natural to expect that the $w$ coefficient
varied in time, i.e. $w = w(z)$. In this paper, we take a Taylor
expansion of $w(z)$ with respect to the scale factor, which leads to
the following redshift dependence: $w(z)=w_0+w_1\frac{z}{1+z}$. This
is so called Chevalier-Polarski-Linder (CPL) model proposed in
\citep{Chevalier01,Linder03}.

It has been known for some time that it is hard to get good and
stringent fits for all parameters in this model. Hence, we fix
matter density parameter at the best-fit value $\Omega_m=0.315$
based on the recent Planck observations \citep{Ade14}. Our best fit
values for the CPL model parameters are
$\left\{w_0,w_1\right\}=\left\{-0.879^{+0.414}_{-0.431},-0.463^{+2.108}_{-1.717}\right\}$,
$\left\{w_0,w_1\right\}=\left\{-0.951^{+0.249}_{-0.247},-0.4034^{+1.160}_{-1.139}\right\}$,
and
$\left\{w_0,w_1\right\}=\left\{-0.965^{+0.154}_{-0.155},-0.241^{+0.498}_{-0.501}\right\}$
from SL+BAO, SN, and SL+BAO+SN, respectively. Confidence regions
(corresponding to 68.3\% and 95.8\% confidence levels) for standard
rulers, standard candles and the combined analysis in the ($w_0$,
$w_1$) plane are shown in Fig.~\ref{tab:CPL}. One can notice that
confidence contours from standard rulers and standard candles are
inclined with respect to each other. This is a promising signal in
light of the colinearity of $w_0$ and $w_1$ parameters (due to their
fundamental anticorrelation), which has so far been a major
obstruction in getting stronger constraint on them. Such inclination
gives us hope that combined analysis will eventually lead to more
precise assessment of $w_0$ and $w_1$ and decide whether cosmic
equation of state evolved or not.

The results obtained with standard rulers turned out to correspond
well with previous work by \citet{Marek11}, whose results (for
standard rulers) were $w_0=-0.993\pm0.207,w_1=0.609\pm1.071$. As far
as standard candles are concerned, the result of joint analysis from
WMAP+BAO+$H_0$+SN given by \citet{Komatsu11} is
$w_0=-0.93\pm0.13,w_1=0.41^{+0.72}_{-0.71}$. Moreover, the combined
analysis of standard rulers and candles performed in \citet{Marek11}
resulted in the following best fits:
$w_0=-0.989\pm0.124,w_1=0.082\pm0.621$. These results are in
agreement with ours within $1\sigma$. In addition, our joint
analysis tends to support the models with a varying equation of
state which are very close to the $\Lambda$CDM model ($w_0=-1$,
$w_1=0$).

\subsection{Ricci Dark Energy model}

There are other cosmological models that have gained a lot
attention, one of them is the so called holographic dark energy,
which is inspired by the holographic principle resulting from
quantum gravity. It is well known that gravitational entropy of a
given closed system with a characteristic length scale $L$ is not
proportional to its volume $L^3$, but to its surface area $L^2$
\citep{Bekenstein81,Gao09}. Because the cosmological constant
$\Lambda$ scales like inverse length squared, one could postulate
this length scale coinciding with present Hubble horizon. This way
the coincidence and fine tuning problem could be alleviated. However
it turned out that such model has troubles with explaining
accelerated expansion of the Universe. Therefore \citet{Gao09}
proposed to choose $|R|^{-1/2}$ as the infrared cutoff, where
$R=-6(\dot{H}+2H^2)$ is Ricci scalar. In this section we will
confront this model with standard rulers and standard candles data.
The density of dark energy in this model is
\begin{equation}
 {\rho_{de}}={3{\beta}M^2_{Pl}(\dot{H}+2H^2)}
\end{equation}
where $\beta>0$ is some constant parameter to be fitted. 

With the methods described above, we obtain the following best fits
for RDE model:
$\left\{\Omega_m,\beta\right\}=\left\{0.201^{+0.017}_{-0.019},0.566^{+0.087}_{-0.086}\right\}$,
$\left\{\Omega_m,\beta\right\}=\left\{0.202^{+0.086}_{-0.088},0.534^{+0.104}_{-0.105}\right\}$,
and
$\left\{\Omega_m,\beta\right\}=\left\{0.201^{+0.016}_{-0.016},0.531^{+0.041}_{-0.041}\right\}$
from SL+BAO, SN, and SL+BAO+SN, respectively. These results are also
illustrated in Fig.~\ref{tab:RDE}. These results are in agreement
with previous work of \citet{Cao12c}. Additionally the best fits for
different probes are very close to each other.

\subsection{Dvali-Gabadadze-Porrati model}

Cosmological models we investigated so far were based on the
Einstein's theory of gravity. There are also other approaches which
seek an explanation of accelerated expansion of the Universe in
modifications of General Relativity. Dvali-Gabadadze-Porrati (DGP)
brane world model is a well-known example of this class, based on
the assumption that our 4-dimensional spacetime is embedded into a
higher dimensional bulk spacetime \citep{Dvali00}.

In this model, the Friedman equation is modified to
\begin{equation} \label{Friedman DGP}
 {H^2+\frac{k}{a^2}}={\left[\sqrt{\frac{\rho}{3M^2_{Pl}}+\frac{1}{4r_c}}+\frac{1}{2r_c}\right]^2}
\end{equation}
where $M_{Pl}=\sqrt{\frac{\hbar c}{8 \pi G}}$ is the (reduced)
Planck mass,  $r_c = \frac{M_{Pl}^2}{2 M_5^2}$ (with $M_5$ denoting
5-dimensional reduced Planck mass) is the crossover scale.
Introducing the omega parameter: $\Omega_{rc}=1/(4r^2_cH^2_0)$, one
can rewrite the Eq.~(\ref{Friedman DGP}) in the form leading to the
Hubble function given in Table~1 (assuming also flat Universe).
Friedman equation leads also to the normalization condition:
$\Omega_k + ( \sqrt{\Omega_{r_c}} + \sqrt{\Omega_m + \Omega_{r_c}}
)^2 = 1$, which simplifies to $\Omega_{r_c} = \frac{1}{4} ( 1 -
\Omega_m)^2$ under assumption of flat Universe. Therefore, the flat
DGP model contains only one free parameter, $\Omega_m$. The best fit
values for the mass density parameter in DGP model obtained from
SL+BAO, SN, and SL+BAO+SN are: ${\Omega_m}=0.269^{+0.014}_{-0.016}$,
${\Omega_m}=0.165^{+0.036}_{-0.035}$, and
${\Omega_m}=0.257^{+0.011}_{-0.012}$, respectively. These results
are also illustrated in Fig.~\ref{tab:DGP}.

Former work done by \citet{Xu10} indicated
${\Omega_m}=0.266^{+0.0298}_{-0.0304}$ and the results of
\citet{Marek11} ${\Omega_m}=0.267\pm0.013$ from CMB+BAO+SL+SN match
accurately with our results. Moreover, our results are in a very
good agreement with \citet{Cao12c}.

Closing this section, let us stress that we have not only
constrained cosmological models, but also we considered the
evolution of slope factor in the mass density profile of lensing
galaxies. In consequence, we also obtained the best fits for
$\gamma$ parameters, as shown in Table~1. One can see, that the
$\gamma$ parameters estimated in different cosmological models are
very similar. It suggests that the method of using distance ratios
from strong lensing systems can be effective in cosmological
applications. More precisely, since the $\gamma$ parameters of lens
mass distribution model seem to be unaffected by cosmological model
assumed, one can hope to calibrate them within say $\Lambda$CDM
model and then use the best fits as an input for cosmological model
testing with the samples like ours (118 lenses) or similar ones
obtained in the future.

\section{Model selection} \label{AIC}

In the previous section, we obtained the best fits for five
cosmological models from 118 galactic-scale strong lensing systems
combined with 6 BAO observations. However, the $\chi^2$ statistic
alone does not provide any way to compare the competing models and
decide which one is preferred by the data. This question can be
answered with model selection techniques \citep{Cao12d}.

Therefore, we used of two criteria: Akaike Criterion (AIC)
\citep{Akaike74} and Bayesian Information Criterion (BIC)
\citep{Schwarz78}. They have became standard in applied statistics,
were first used in cosmology by \citet{Liddle} and then e.g. by
\citet{Szydlowski} or \citet{BiesiadaAIC}. The value of AIC --
approximately unbiased estimator of the Kullback-Leibler divergence
between the given model and the ``true'' one can be calculated as
 \begin{equation}
 AIC=-2lnL_{max}+2k
 \end{equation}
where $L_{max}$ is the maximum likelihood value, $k$ is the number
of free parameters in the model. In our case $k$ comprises both
cosmological parameters and galaxy mass density slopes. If the
uncertainties are Gaussian likelihood can be calculated from the
chi-square function $\chi^2=-2lnL_{max}$. The value of $AIC$ for a
single model is meaningless. What is useful is the difference in AIC
between cosmological models $\Delta AIC$. This difference is usually
calculated with respect to the model which has the smallest value of
AIC
 \begin{equation}
 \Delta AIC(i)=AIC(i)-AIC_{min}
 \end{equation}
where the index $i=1,...,5$ represents cosmological models and
$AIC_{min}=min\left\{AIC(i)\right\}$. BIC is defined in a very
similar manner to AIC, but it adds the information about the sample
size $N$:
 \begin{equation}
 BIC={\chi^2}+2k ln N
 \end{equation}

For the purpose of model selection we only used the standard rulers,
i.e. BAO combined with 118 lensing data. Table 3 lists the AIC and
BIC difference of each model. One can see that both $AIC$ and $BIC$
criteria support $\Lambda$CDM as the best cosmological model, in the
light of current observational strong lensing data. Concerning the
ranking of other competing models, AIC and BIC criteria give
different conclusions. According to the AIC, next are the RDE and
XCDM model: odds against them with respect to $\Lambda$CDM (see
\citep{BiesiadaAIC} for details) are 2.6:1 (they differ at the
second decimal place). Then the CPL and DGP are slightly worse
supported, with odds against 2.8:1. In summary, one can say that
besides the $\Lambda$CDM as the best one, all other models get
similar support by standard rulers. On the other hand BIC criterion
gives a different ranking: next after $\Lambda$CDM is the DGP brane
model with odds against equal to 2.8:1. Then there are RDE (odds
against 10.1 : 1) and XCDM (odds against 10.5:1) while CPL model
gets the smallest support with odds against 11.5:1. In summary, one
can state that BIC substantially penalizes cosmological models with
more than one free parameter. In particular our findings are in
contrast with \citep{BiesiadaAIC} who claimed that DGP model is
strongly disfavored by the data.

 \begin{table}
 \begin{center}
 \caption{ Summary of the information criteria, AIC and BIC for the combined SL+BAO data.}
 \begin{tabular}{c|c|c|c|ccc}
 \hline \hline
 Model & AIC  & $\Delta$AIC  & BIC  & $\Delta$BIC\\
 \hline $\Lambda$CDM  & 320.71 & 0  & 323.53 & 0 \\
 \hline XCDM  & 322.61 &  1.89 & 328.25 &  4.71\\
 \hline CPL   & 322.77¡¡&  2.06 & 328.41¡¡&  4.88\\
 \hline RDE & 322.59 & 1.88 & 328.23 & 4.63\\
 \hline DGP &  322.76  &  2.05 &  325.58  &  2.05\\
 \hline \hline
 \end{tabular}
 \end{center}
 \end{table}

\section{Summary and Conclusion} \label{conclusions}

In this paper, we used standard rulers and standard candles
(separately and jointly) to explore five popular dark energy models
under assumption of spatial flatness of the Universe. As standard
rulers, we used new galactic-scale strong lensing data set compiled
by \citet{Cao15} combined with BAO diagnostics. Supernovae Ia served
as standard candles. In order to compare the degree of support given
by the standard rulers to various competing dark energy models, we
performed a model selection using the AIC and BIC information
criteria.

The main conclusions of this paper can be summarized as follows.
Firstly, relaxing the mass density profile of SIS model to a more
general power-law density profile, the best fitted $\gamma$
parameters are in agreements with each other irrespective of the
cosmological model considered. This demonstrates that inclusion of
mass density power index as a free parameter does not lead to
noticeable spurious effects of mixing them with cosmological
parameters in statistical procedure of fitting. Therefore, we can
say that galactic strong lensing systems may provide a complementary
probe to test the properties of dark energy. Secondly, because
standard rulers and standard candles have different parameter
degeneracies in cosmology, joint analysis of standard rulers and
standard candles gives much more restrictive results for
cosmological parameters. Thirdly, the information theoretic criteria
(AIC and BIC) support the claim, that cosmological constant model is
still the best one and there is no reason to prefer any more complex
model. In light of forthcoming new generation of sky surveys like
the EUCLID mission, Pan-STARRS, LSST, JDEM, which are estimated to
discover from thousands to tens of thousands of strong lensing
systems it would be interesting to stay tuned and look forward to
seeing whether this conclusion could be changed by more
gravitational lensing systems observed in the future.


\begin{acknowledgements}
This work was supported by the Ministry of Science and Technology
National Basic Science Program (Project 973) under Grants Nos.
2012CB821804 and 2014CB845806, the Strategic Priority Research
Program "The Emergence of Cosmological Structure" of the Chinese
Academy of Sciences (No. XDB09000000), the National Natural Science
Foundation of China under Grants Nos. 11503001, 11373014 and
11073005, the Fundamental Research Funds for the Central
Universities and Scientific Research Foundation of Beijing Normal
University, and China Postdoctoral Science Foundation under grant
No. 2014M550642 and 2015T80052. Part of the research was conducted
within the scope of the HECOLS International Associated Laboratory,
supported in part by the Polish NCN grant DEC-2013/08/M/ST9/00664 -
M.B. gratefully acknowledges this support. This research was also
partly supported by the Poland-China Scientific \& Technological
Cooperation Committee Project No. 35-4. M.B. obtained approval of
foreign talent introducing project in China and gained special fund
support of foreign knowledge introducing project. He also gratefully
acknowledges hospitality of Beijing Normal University where this
project was initiated and developed.
\end{acknowledgements}

\end{document}